\documentclass[trackchanges,twocolumn]{aastex701}

\usepackage{booktabs}

\newcommand{\psec}{s$^{-1}$}  
\newcommand{\erg}{erg cm$^{-2}$ s$^{-1}$} 
\newcommand{\lum}{erg s$^{-1}$} 
\newcommand{\lxp}{\emph{AstroSat}/LAXPC}
\newcommand{\sxt}{\emph{AstroSat}/SXT}

\def \src{{SAX J1808.4--3658}}
\def \nustar{\emph{NuSTAR}}
\def \nicer{\emph{NICER}}
\def \astrosat{\emph{AstroSat}}

\begin{document}

\title{Broadband Timing and Spectral Study of Accreting Millisecond X-ray Pulsar \src\ during Its 2022 Outburst}

\author[orcid=0000-0003-0366-047X]{Rahul Sharma}
\email[show]{rahul1607kumar@gmail.com}
\affiliation{Inter-University Centre for Astronomy and Astrophysics (IUCAA), Ganeshkhind, Pune 411007, India}

\author[orcid=0000-0002-0118-2649]{Andrea Sanna}
\email{}
\affiliation{Universit\'{a} degli Studi di Cagliari, Dipartimento di Fisica, SP Monserrato-Sestu, KM 0.7, 09042 Monserrato, Italy}

\author[orcid=0000-0003-4696-940X]{Prince Sharma}
\email{}
\affiliation{Department of Applied Sciences, UPES, Bidholi, Dehradun, Uttarakhand 248007, India}

\begin{abstract}
We report on our investigation of the \nustar\ and \astrosat\ observations along with simultaneous \nicer\ observations of the accreting millisecond X-ray pulsar \src, obtained during its tenth outburst from 2022. The \nustar\ observation captured the source near the outburst peak, while \astrosat\ observed it during the decay phase. Coherent pulsations at $\sim$401 Hz were detected throughout the outburst, with the fundamental amplitude in the 3--30 keV range increasing from $\sim$4\% near the peak to $\sim$6\% during the decay. The pulsations display strong energy dependence and negative time lags of $\sim$0.2--0.3 ms, with harder photons leading softer ones. The broadband spectra in both epochs are well described by a soft thermal component and Comptonized continuum, together with a prominent relativistic reflection component. As the outburst evolved, the continuum softened ($\Gamma$ increasing from $\sim$1.88 to $\sim$1.99) and the coronal electron temperature decreased ($kT_{\rm e}$ from $\sim$31 to $\sim$18 keV), consistent with enhanced Compton cooling at lower accretion rates. The ionization parameter declined ($\log \xi$ from $\sim$3.4 to $\sim$1.8) while the reflection fraction increased, suggesting a changing accretion geometry with a more compact corona and a larger disk covering fraction during the decay phase. The X-ray luminosity decreased by a factor of $\sim$3 between the two epochs. Our results suggest the coupled evolution of the corona, disk, and magnetosphere as the mass accretion rate declines. 
\end{abstract}



\section{Introduction}
\label{intro}

Accretion-powered millisecond X-ray pulsars (AMXPs) form a special class of neutron star low-mass X-ray binaries (NS-LMXBs) where an old NS in a compact binary is accelerated to very high spin due to accretion from its companion. The spin period in such systems is observed to be in the range of milliseconds \citep{Srinivasan2010}. \src\ is the first LMXB discovered to show millisecond X-ray pulsations with a spin frequency of $\sim$401 Hz in a compact binary with an orbital period of $\sim$2 hr \citep{Wijnands1998, Chakrabarty1998}. Currently, a total of 24 sources have been identified and cataloged that shows millisecond pulsations \citep[e.g.,][]{Ng2021, Patruno2021, Bult2022, DiSalvo2022, Sanna2022b}.

Several studies suggest that the magnetic field of AMXPs ranges between $\sim$$10^{8}$ and $10^{9}$ G \citep[see, e.g.,][]{Mukherjee2015, Ludlam2017c, Sharma2020}. Although relatively weak, such a field proves strong enough to channel the accreting material onto the magnetic poles, resulting in X-ray modulation at the spin frequency. Spectral properties of AMXPs typically show the characteristics of thermal emission from the accretion disk and/or NS surface, thermal Comptonization, and reprocessed emission from the accretion disk in the form of a reflection spectrum \citep[e.g.,][]{Papitto2009, Cackett2010, Sharma2019, DiSalvo2022}. These systems are often observed in the hard spectral state, although some studies caught these sources in transition states between the hard and soft states \citep{Sharma2019, DiSalvo2022, Beri2023}, similar to those seen in atoll sources \citep[e.g.,][]{Hasinger}.

Since its discovery, \src\ has exhibited recurrent outbursts every 2.5--4 yr. To date, 11 such events have been recorded: in 1996, 1998, 2000, 2002, 2005, 2008, 2011, 2015, 2019, 2022, and 2025, the largest number of known outbursts for any AMXP \citep{DiSalvo2022, Illiano23, Russell25ATel}. Owing to this frequent activity, \src\ has been the subject of extensive multiwavelength investigations, spanning quiescence to outburst phases across radio, optical, UV, and X-ray bands \citep[e.g.,][]{Patruno2017, Baglio2020, Goodwin2020, Ambrosino2021, Sharma2023-SAX}. 

This work focuses on the 2022 outburst of \src, which began around 2022 August 19 \citep{Imai2022ATel, Sanna2022ATel}. About five days after its onset, the flux started to decline, and the source subsequently entered the characteristic reflaring stage that marks the late phases of its outbursts, which lasted for more than a month \citep{Baglio2022ATel, Illiano2022ATel}. At peak, the 0.6--10 keV luminosity reached $\simeq1\times10^{36}$ \lum. The 2022 event displayed a somewhat unusual outburst profile compared to the typical behavior of \src: the normally short-lived peak persisted for the longest duration recorded so far, while the slow-decay/rapid-drop phase was noticeably shorter ($\sim$10--15 days) than usual \citep{Illiano23}. The Neutron Star Interior Composition Explorer \citep[\emph{NICER};][]{nicer} tracked the outburst from its onset through the reflaring phase. Using these NICER data, \citet{Illiano23} carried out a coherent timing analysis, reporting a secular spin-down of the pulsar at a rate of $1.15(6)\times10^{-15}$ Hz s$^{-1}$. They also found evidence for a decrease in the orbital period, with the orbital phase epochs exhibiting an $\sim$11 s modulation consistent with a $\sim$21 yr periodicity. 

In addition to the \nicer\ monitoring, the 2022 outburst was also followed up with Nuclear Spectroscopic Telescope ARray (\nustar) and \astrosat\ near the peak and decay phase of the outburst, respectively. Recently, \cite{Kaushik25} reported the evolution of spectral and aperiodic timing properties using \nicer\ and \astrosat. \cite{Bruce26} reported a broadband spectral study with \nicer+\nustar\ observations comparing the 2019 and 2022 outbursts.
In this work, we utilized the \nicer, \nustar, and \astrosat\ observations to investigate the pulse timing and spectral properties of \src\ across a broad energy range during the peak and decay phase of the 2022 outburst. The structure of the paper is as follows: Section \ref{obs} describes the data reduction and analysis techniques. The timing and spectral analysis results are summarized in Section \ref{res}. 
We discuss our findings in Section \ref{dis} and conclude in Section \ref{summary}.

\section{Observation and data analysis}
\label{obs}

\begin{figure*}
\centering
 \includegraphics[width=0.9\linewidth]{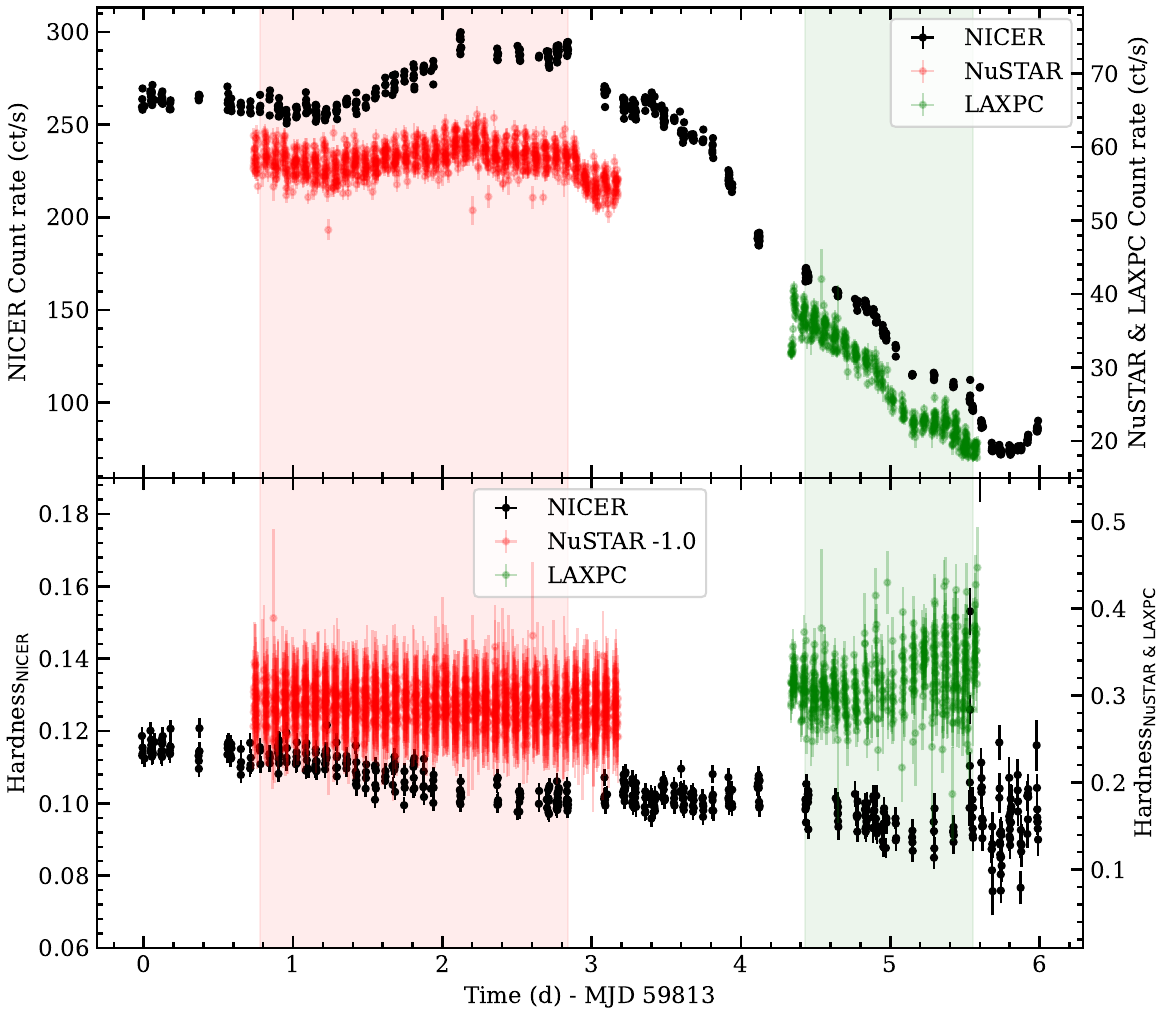}
 \caption{The light curve of \src\ during its 2022 outburst obtained with \nicer, \nustar\ and \lxp\ binned at 100 s in the energy ranges 0.5--10 keV, 3--30 keV and 3--30 keV, respectively. The lower panel shows the hardness ratio, defined as the count-rate ratios between 3–10 and 0.5–3 keV for \nicer, and between 10–30 and 3–10 keV for \nustar\ and \lxp. The hardness values for \nustar\ have been shifted by $-$1 to match LAXPC. The shaded region marks the NICER segment that was combined for further analysis.} 
\label{fig:lc}
\end{figure*}

\begin{table*}
\caption{Log of X-ray observations of \src\ analyzed in this work.}
\centering
\resizebox{\linewidth}{!}{
\hskip-2.2cm\begin{tabular}{c c c c c c c c c}
\hline
Epoch & Instrument & obsID & \multicolumn{2}{c}{Start Time} & \multicolumn{2}{c}{Stop Time} & Exposure\\
\cmidrule(lr){4-5}\cmidrule(lr){6-7}
&&& (yy-mm-dd hh:mm:ss) & MJD & (yy-mm-dd hh:mm:ss) & MJD & (ks) \\
\hline
1 & \nustar & 80701312002 & 2022-08-22 17:36:31 & 59813.73369 & 2022-08-25 04:16:06 & 59816.17785 & 107 \\
& \nicer --1 & $a$ & 2022-08-22 18:38:10 & 59813.77650 & 2022-08-24 20:14:26 & 59815.84335 & 20 \\
\hline
2 & \lxp\ & 9000005318 & 2022-08-26 08:00:39 & 59817.33379 & 2022-08-27 13:59:59 & 59818.58332 & 50 \\
& \sxt\ & 9000005318 & 2022-08-26 08:24:57 & 59817.35066 & 2022-08-27 13:55:01 & 59818.57988 & 27 \\
& \nicer --2 & $b$ & 2022-08-26 10:20:36 & 59817.43097 & 2022-08-27 13:20:26 & 59818.55585 & 9.4 \\
\hline
\multicolumn{7}{l}{$^a$ Observation covers snapshots from obsIDs: 5050260104, 5050260105, 5574010101, 5574010102, and 5050260106.}\\
\multicolumn{7}{l}{$^b$ Observation covers snapshots from obsIDs: 5574010104 and 5574010105.}\\
\end{tabular}}
\label{tab:obslog}
\end{table*}

The observations used in this work are detailed in Table \ref{tab:obslog}.

\subsection{NuSTAR}

The \nustar\ \citep{Harrison13} mission comprises two telescopes that focus X-rays between 3 and 78 keV onto two identical focal planes (usually called focal plane modules A and B, or FPMA and FPMB). 

For this work, we have used the most recent \nustar\ analysis software distributed with HEASOFT version 6.34 and the latest calibration files (v20250122) for reduction and analysis of the \nustar\ data. The calibrated and screened event files have been generated by using the task \textsc{nupipeline}. A circular region of radius $70^{\prime\prime}$ centred at the source position was used to extract the source events. Background events were extracted from a circular region of the same size away from the source. The task \textsc{nuproduct} was used to generate the light curves, spectra, and response files. The FPMA and FPMB light curves were background-corrected and summed using \textsc{lcmath}. The spectra were grouped using \textsc{ftgrouppha} by following the \citet{Kaastra} optimal binning scheme with a minimum of 25 counts per bin.

\subsection{AstroSat}

\astrosat\ is India's first dedicated multiwavelength astronomy satellite \citep{Agrawal2006, Singh2014}, launched in 2015. It has five principal payloads on board: (i) the Soft X-ray Telescope (SXT), (ii) the Large Area X-ray Proportional Counters (LAXPCs), (iii) the Cadmium-Zinc-Telluride Imager (CZTI), (iv) the Ultra-Violet Imaging Telescope (UVIT), and (v) the Scanning Sky Monitor (SSM). In this work, we have only analyzed data from SXT and LAXPC.

\subsubsection{LAXPC}

LAXPC is one of the primary instruments aboard \astrosat. It consists of three coaligned identical proportional counters (LAXPC10, LAXPC20, and LAXPC30) that work in the energy range 3--80 keV. Each LAXPC detector independently records the arrival time of each photon with a time resolution of $10 ~\mu$s and has five layers \citep[for details see][]{Yadav2016, Antia2017}.

As LAXPC10 was operating at low gain and detector LAXPC30\footnote{LAXPC30 has been switched off since 2018 March 8 due to abnormal gain changes; see \url{http://astrosat-ssc.iucaa.in/}} was switched off during the observation, we used only the LAXPC20 detector for our analysis. We used the data collected in the event analysis (EA) mode and processed them using the \textsc{LaxpcSoft}\footnote{\url{http://www.tifr.res.in/~astrosat\_laxpc/LaxpcSoft.html}} v3.4.4 software package to extract light curves and spectra. 
LAXPC detectors have a dead-time of $42~\mu$s, and the extracted products are dead-time corrected.
The background in LAXPC is estimated from the blank-sky observations \citep[see][for details]{Antia2017}. To minimize the background, we have performed all analyses using the data from the top layer (L1, L2) of the LAXPC20 detector. We have used corresponding response files to obtain channel-to-energy conversion information while performing energy-resolved analysis.

\subsubsection{SXT}

SXT is a focusing X-ray telescope with CCD in the focal plane that can perform X-ray imaging and spectroscopy in the 0.3--7 keV energy range \citep{Singh2016, Singh2017}. 
\src\ was observed in the photon counting (PC) mode with SXT. Level 1 data were processed with \texttt{AS1SXTLevel2-1.4b} pipeline software to produce level 2 clean event files. The level 2 cleaned files from individual orbits were merged using the SXT event merger tool\footnote{\url{https://github.com/gulabd/SXTMerger.jl}}. The merged event file was then used to extract images and spectra using the \textsc{xselect} task, provided as part of \textsc{heasoft} v6.31.1. A circular region with a radius of $16^{\prime}$ centred on the source was used.
No source pileup was observed as the count rate was below the threshold limit for pileup ($<40$ counts \psec) in the PC mode\footnote{\url{https://www.tifr.res.in/~astrosat_sxt/instrument.html}}. For spectral analysis, we have used the blank-sky SXT spectrum as background (SkyBkg\_sxt\_LE0p35\_R16p0\_v05\_Gd0to12.pha) and spectral redistribution matrix file (sxt\_pc\_mat\_g0to12.rmf) provided by the SXT team\footnote{\url{http://www.tifr.res.in/~astrosat\_sxt/dataanalysis.html}}. We generated the correct off-axis auxiliary response files (ARF) using the sxtARFModule tool from the on-axis ARF (sxt\_pc\_excl00\_v04\_20190608.arf) provided by the SXT instrument team. 

\subsection{NICER}

\nicer\ \citep{nicer} is an X-ray telescope deployed on the International Space Station (ISS) in 2017 June. \nicer\ X-ray Timing Instrument has 56 aligned FPMs, each made up of an X-ray concentrator optic associated with a silicon drift detector. It has a large effective area and high temporal resolution in the soft X-ray band. Each NICER observation typically comprises multiple short-duration pointings of the source, called snapshots, primarily driven by ISS orbit and visibility constraints.

\nicer\ monitored \src\ after the onset of the outburst from MJD 59810 to 59883. \nicer\ data were processed with \textsc{heasoft} v6.34 and the \nicer\ Data Analysis Software (\texttt{nicerdas}) v2024-08-18\_V013, using Calibration Database (CALDB) xti20240206. Standard calibration and screening criteria were applied using the \texttt{nicerl2} tool. The \nicer\ campaign overlapped quasi-simultaneously with the \nustar\ and \astrosat\ observations. Since individual \nicer\ observations typically spanned less than a day, we merged segments overlapping with \nustar\ and \astrosat\ using \texttt{niobsmerge} (Fig. \ref{fig:lc}). The \nicer-1 contains snapshots of obsIDs 5050260104, 5050260105, 5574010101, 5574010102, and 5050260106, while \nicer-2 covers obsIDs 5574010104 and 5574010105 (see Table \ref{tab:obslog}).
The merged event files were then used to extract light curves with \texttt{nicerl3-lc} and spectra with \texttt{nicerl3-spect}. Background contributions were estimated using the \texttt{nibackgen3C50} tool \citep{Remillard22}.

All photon arrival times were corrected to the solar system barycentre using \textsc{as1bary}\footnote{\url{http://astrosat-ssc.iucaa.in/?q=data\_and\_analysis}} for \lxp\ and \textsc{barycorr} for \nustar\ and \nicer, adopting the JPL DE405 planetary ephemeris. For \nustar, we used the latest clock correction file available with the calibration database v20250122. The source position used for the corrections was R.A. (J2000) = $18^{\mathrm h} 08^{\mathrm m} 27^{\mathrm s}.647$ and decl. (J2000) = $-36^{\circ} 58' 43.90''$ \citep{Bult2020}.

\section{Results}
\label{res}

\subsection{Light Curve}

Fig. \ref{fig:lc} shows the light curve of \src\ during its 2022 outburst as obtained with \nicer, \nustar\ and \lxp, binned at 100 s in the energy ranges 0.5--10 keV, 3--30 keV and 3--30 keV, respectively. The \nustar\ observation was carried out near the peak of the outburst, while the \astrosat\ observation took place during the decay phase. The shaded region marks the \nicer\ segment that was merged with respect to \nustar\ and \astrosat\ observations. These two observations are hereafter referred to as Epochs 1 and 2, respectively. 
The bottom panel of Fig. \ref{fig:lc} shows the hardness ratio computed from light curves in the two energy bands 3--10 and 0.5--3 keV for \nicer, and between 10--30 and 3--10 keV for \nustar\ and \lxp. During the \nicer\ monitoring, the hardness ratio exhibits a slight decreasing trend over time. A more pronounced change is observed between the two epochs, with the hardness ratio decreasing from $\sim$1.3 during the \nustar\ observation to $\sim$0.3 during the \astrosat\ observation, indicating significant spectral softening as the outburst evolved. Although the \astrosat\ and \nicer-2 observation shows clear flux evolution, no significant hardness variations were detected in either epoch, suggesting that the source did not undergo a spectral state transition within these intervals.

\begin{table}
\caption{Spin frequency and fractional amplitude.}
\centering
\resizebox{\columnwidth}{!}{
\hskip-1.0cm\begin{tabular}{llll}
\hline 
Instrument & Spin Frequency (Hz) & \multicolumn{2}{c}{Fractional Amplitude (\%)} \\
\cmidrule(lr){3-4}
 & & Fundamental & Harmonic \\
\hline
\nustar & 400.9752098 (1) & 4.06 (11) &  1.71 (11) \\
\nicer-1 & 400.9752096 (1) & 4.44 (6) & 0.80 (6) \\
    \hline
\astrosat & 400.9752096 (2) & 6.32 (16) &  1.72 (16) \\
\nicer-2 & 400.9752089 (3) & 6.54 (7) & 0.26 (7) \\
    \hline
\end{tabular}}
\label{tab:spin}
\end{table}

\begin{figure*}
\centering
 \includegraphics[width=0.48\linewidth]{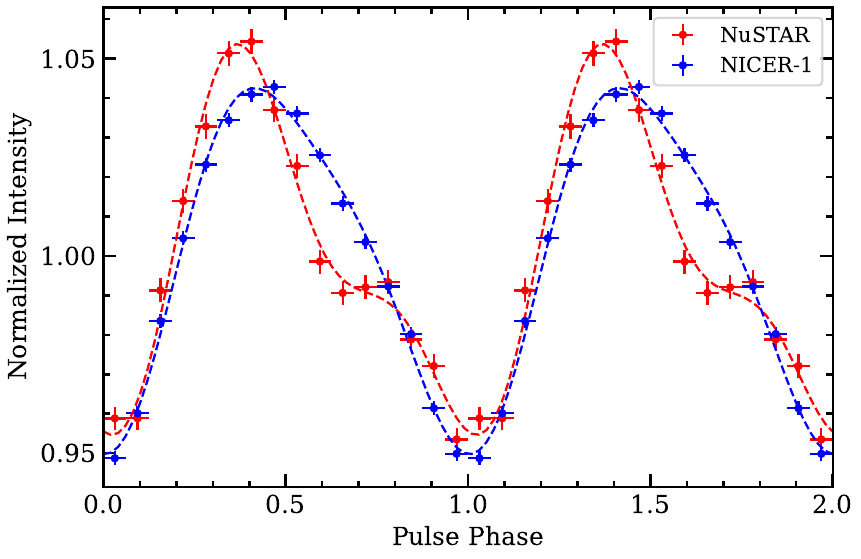}
  \includegraphics[width=0.48\linewidth]{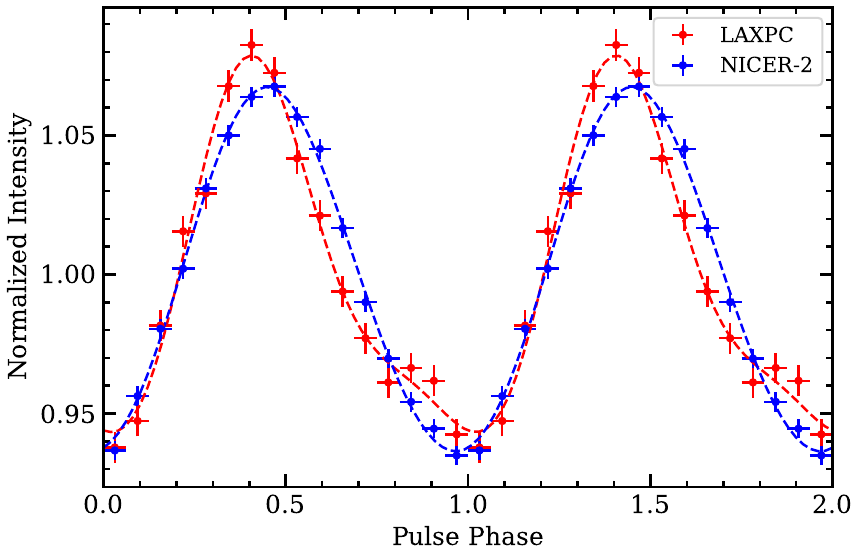}
 \caption{Pulse profiles of \src\ from two observing epochs, obtained by epoch-folding the data with spin frequencies listed in Table \ref{tab:spin} after correcting for the orbital motion. Left panel: Epoch 1, showing profiles in the 0.5--10 keV band from \nicer\ and in the 3--30 keV band from \nustar. Right panel: Epoch 2, showing profiles in the 0.5--10 keV band from \nicer\ and in the 3--30 keV band from \lxp. The solid lines denote the best-fitting model, which is the superposition of two sinusoidal functions with harmonically related periods. For clarity, two pulse cycles are shown. }
\label{fig:pp}
\end{figure*}

\begin{figure*}
\centering
 \includegraphics[width=0.48\linewidth]{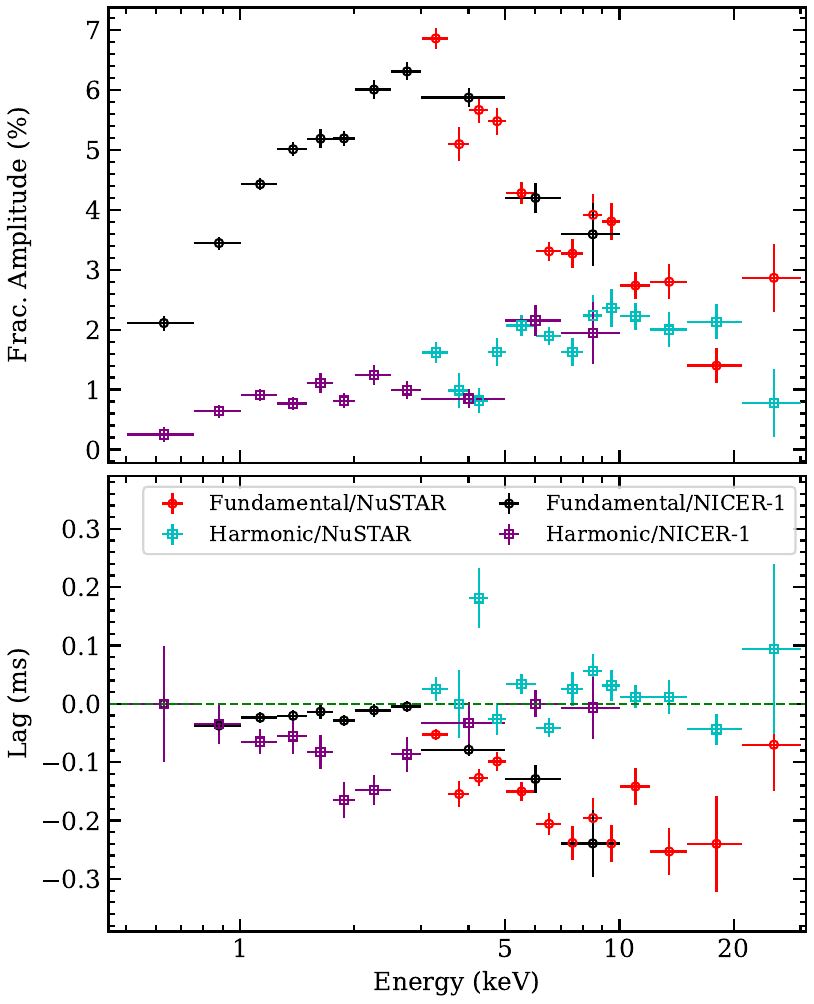}
 \includegraphics[width=0.48\linewidth]{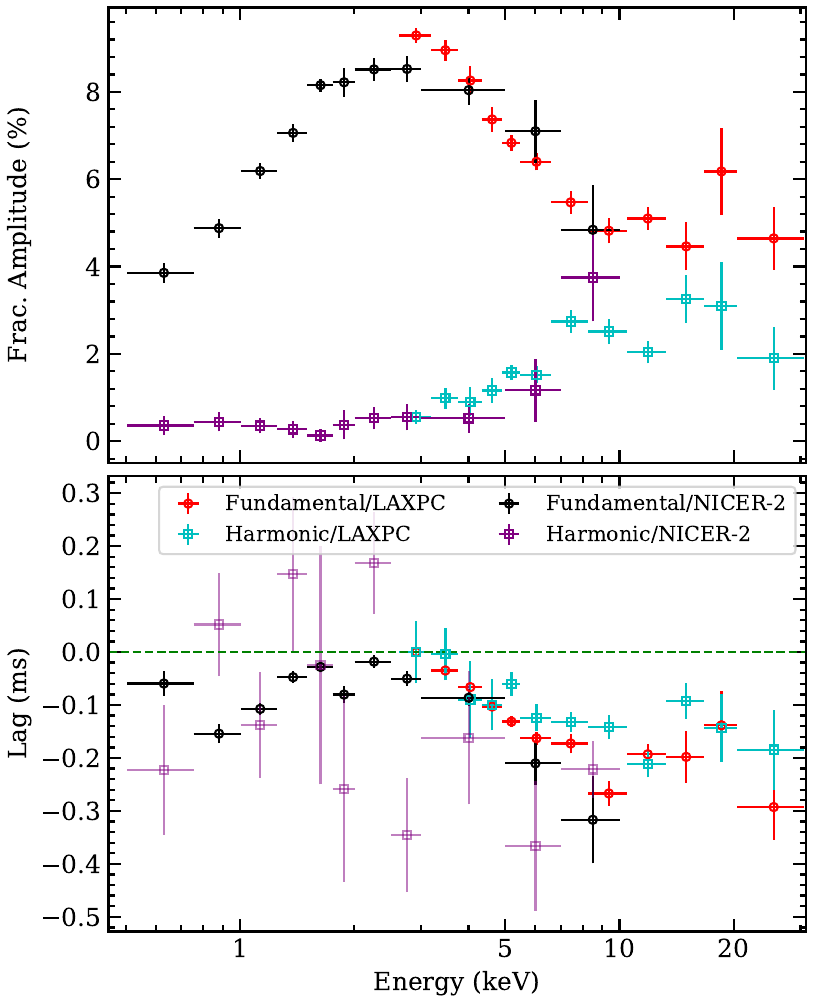}
 \caption{Energy-resolved pulse properties of \src. Top panels: fractional amplitudes of the fundamental and harmonic components as a function of energy. Bottom panels: corresponding time lags of the fundamental and harmonic components. The left plots correspond to Epoch 1, with the zero lag defined relative to the pulse phase at 0.5 keV, while the right plots correspond to Epoch 2, with the zero lag defined at 3 keV.}
\label{fig:energy-res}
\end{figure*}

\subsection{Timing analysis}
\label{timing}

Pulses from a rotating NS can lose coherence on relatively short timescales due to Doppler modulation of the photon arrival times in the short $\sim$2 hr orbit. In addition, the low pulse fraction makes it challenging to detect pulsations. To recover the pulsations and measure the true spin frequency during these observations, the photon arrival times must be corrected for binary motion. The time delay induced by the orbital motion, assuming a circular (or very low eccentricity) orbit, is given by \citep{Burderi2007}
\begin{equation}
\label{eq:bary}
    \frac{z(t)}{c}= \frac{a_x \sin i}{c}\,\sin\Big(\frac{2\pi}{P_{\rm orb}} \,(t-T^\star)\Big),   
\end{equation}
where $a_x \sin i / c$ is the projected semi-major axis of the neutron star orbit in light-seconds, $P_{\rm orb}$ is the orbital period, and $T^\star$ is the time of passage through the ascending node.

The most recent orbital solution, derived from \nicer\ observations during the 2022 outburst, was reported by \citet{Illiano23}. We therefore corrected the photon arrival times of the observations used using this updated ephemeris. To search for pulsations, we applied epoch-folding around the spin frequency $\nu_0 = 400.97520953$ Hz, as measured during the same outburst with \nicer\ monitoring \citep{Illiano23}. The signal was sampled with 16 phase bins, and the frequency space was explored in steps of $10^{-8}$ Hz over 10,001 trials. The most significant pulse profile was obtained for an average local spin frequency ($\bar{\nu}$), which is listed in Table \ref{tab:spin} for each dataset used. The uncertainty on $\bar{\nu}$ was estimated using Monte Carlo simulations with 1000 synthetic realizations of the data. Starting from the observed event list and good time intervals (GTIs), we folded the events at the best frequency to obtain a pulse profile. This profile was then used as a template to generate synthetic event lists that preserve the source variability and observational gaps. Each synthetic event list was subjected to an independent frequency search using the epoch-folding statistic. The resulting distribution of best-fit frequencies from the 1000 trials was used to determine the statistical uncertainty on $\bar{\nu}$.

Since the orbital solution is already well constrained by long-term \nicer\ monitoring, and no spin frequency derivative was detected during the 2022 outburst \citep{Illiano23}, we assumed $\bar{\nu}$ to represent the true spin frequency for each observation. 
We also attempted to track pulse phase delays following the method of \citet{Sharma2023-SAX}, but found no significant evidence for 
deviations from the orbital solution of \citet{Illiano23}.

Fig.~\ref{fig:pp} presents the pulse profile of \src, obtained by epoch-folding the \nicer, \nustar, and \lxp\ datasets at $\bar{\nu}$ into 16 phase bins in the energy ranges 0.5--10 keV, 3--30 keV, and 3--30 keV, respectively. The pulse shape is well described by the sum of two harmonically related sinusoidal functions with background-corrected fractional amplitude of the fundamental and harmonic components listed in Table \ref{tab:spin}. The fractional amplitude for the fundamental increases from $\sim$4\% in Epoch 1 to $\sim$6\% in Epoch 2 across the broad energy range. For the harmonic, the amplitude remains nearly constant in the 3--30 keV band, but decreases in the 0.5--10 keV band, from 0.8\% in Epoch 1 to 0.26\% in Epoch 2. A third harmonic is marginally detected in the 3--30 keV band with a fractional amplitude of $\sim$0.3\%, although its significance is low, with F-test false-alarm probabilities of 0.01 and 0.05 for Epochs 1 and 2, respectively.

We then investigated the energy dependence of the pulse amplitude and phase. The full dataset was divided into different energy bands, and each was epoch-folded at $\bar{\nu}$. The resulting pulse profiles were fitted with two harmonically related sinusoidal functions. Fig. \ref{fig:energy-res} (top panels) shows the variation of background-corrected fractional amplitude of the fundamental and harmonic components, with Epoch 1 on the left and Epoch 2 on the right. In both epochs, the fractional amplitude of the fundamental and harmonic was found to be energy-dependent. The fundamental component showed an increase up to 3 keV and then a decreasing trend at higher energies for both epochs. The peak fractional amplitude of the fundamental reached $\sim$7\% in Epoch 1 and was higher in Epoch 2 at $\sim$9\%. The harmonic component in Epoch 1 increased steadily with energy, reaching $\sim$2\% at the higher energies. In Epoch 2, however, the harmonic was not significantly detected at low energies (below 5 keV, $\sim$0.5\%) but showed a rising trend above 5 keV, reaching $\sim$3\% at higher energies.

The bottom panels of Fig. \ref{fig:energy-res} show the energy dependence of the time lag, derived from the phase lag, for the fundamental and harmonic components. In Epoch 1, the fundamental component exhibited no significant lag below 3 keV, while at higher energies it showed a negative lag, reaching $\sim$0.25 ms and indicating that the hard photons arrive earlier than the soft photons. The harmonic component in this epoch showed no clear lag, except for a negative lag around 2 keV. In Epoch 2, the fundamental component again displayed a negative lag above 3 keV, consistent with Epoch 1. However, unlike Epoch 1, the harmonic component in Epoch 2 exhibited a negative lag of $\sim$0.2 ms above 3 keV, while no significant trend was observed below 3 keV owing to its low detection significance.

\subsection{Spectral analysis}
\label{spectral}
 
For both epochs, \nustar\ (FPMA and FPMB) and \astrosat\ (SXT and LAXPC) were fitted with the simultaneous \nicer\ observations. We performed spectral fitting using \textsc{xspec} v12.14.1 \citep{Arnaud} with the component \texttt{tbabs} to model interstellar neutral hydrogen absorption \citep{Wilms}. We have added a constant to account for the cross-calibration between multiple instruments during simultaneous fitting. As the \nicer\ spectra showed some systematic emission feature at 1 keV, which was not observed with SXT, we restricted the \nicer\ spectral fitting to 1.2--10 keV. We also applied a gain correction for the SXT. The gain slope was fixed to 1.0, and the offset was allowed to vary. The gain offset was found to be $\sim$24 eV. A systematic uncertainty of 1\% was added to SXT and LAXPC spectra \citep{Antia2017}. All uncertainties and limits in this paper for spectral analysis correspond to a 90\% confidence level. 

\begin{figure}
\centering
 \includegraphics[width=\columnwidth]{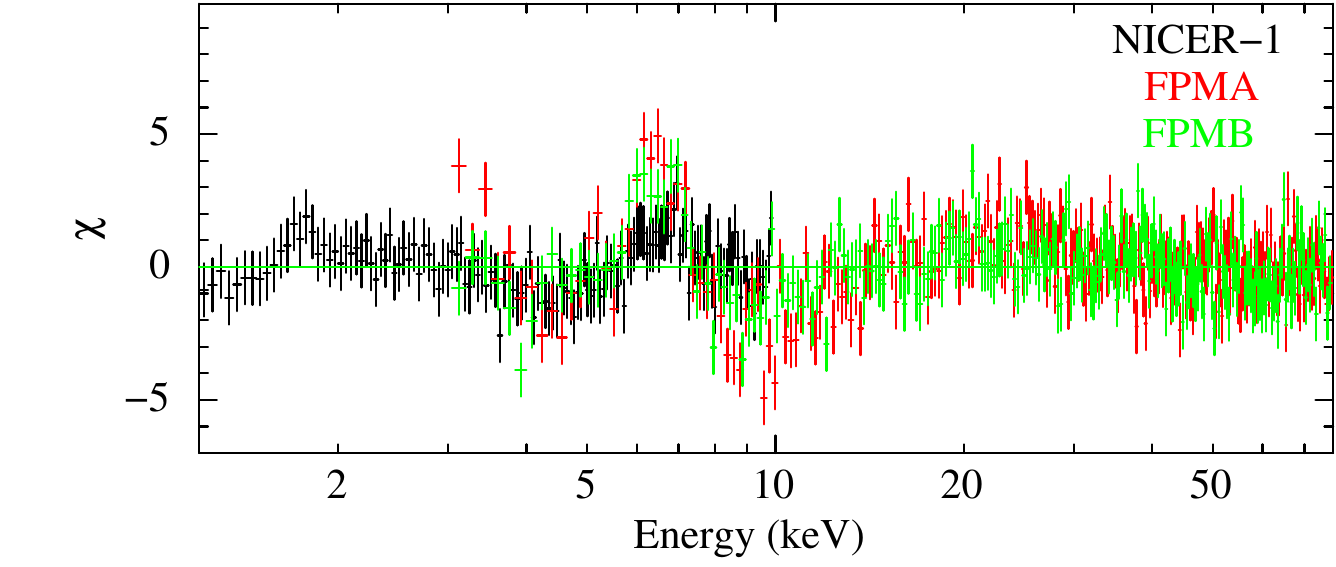}
 \caption{Residuals obtained after modeling the spectrum with absorbed blackbody and Comptonization model. These residuals indicate the broad emission feature at 6--7 keV.}
\label{fig:res}
\end{figure}

The broadband continuum of Epoch 1 can be well described by a combination of a soft thermal component and thermal Comptonization. We modeled the soft emission with the \texttt{bbodyrad} model in \textsc{xspec} and the Comptonized continuum with \texttt{nthcomp} \citep{Zdziarski}. This model, however, provides a statistically unacceptable fit ($\chi^2$/dof = 1035/614) and exhibited prominent residuals around 6--7 keV (Fig. \ref{fig:res}), indicative of Fe K emission \citep[e.g.,][]{Cackett2009, Papitto2009, Salvo2019}, as well as a residual near 1.7 keV in the \nicer\ data, likely associated with Si K calibration uncertainties. Adding a Gaussian emission line significantly improved the fit ($\chi^2$/dof = 603/611), yielding a centroid energy of $6.2 \pm 0.2$ keV, a broad line width of $1.0 \pm 0.1$ keV, and an equivalent width of $0.16 \pm 0.03$ keV. Including a second Gaussian component at 1.7 keV led to only a marginal improvement ($\Delta\chi^2 \sim 8$ for three additional degrees of freedom) with an equivalent width of $\sim$15 eV, and it was therefore excluded from subsequent modeling. The best-fit continuum corresponds to a blackbody temperature of $\sim$0.64 keV and a Comptonizing corona with photon index of $\Gamma \sim 1.86$, electron temperature $kT_e\sim 27$ keV, and seed photon temperature of $\sim$0.2 keV, likely originating from the inner accretion disk.

The presence of the broad emission line (see Fig. \ref{fig:res}) motivated the inclusion of a self-consistent reflection model \citep{Fabian10}. We therefore employed the \texttt{relxillCP} model \citep{Garcia18} of the \texttt{relxill} package v2.4 \citep{Dauser14, Garcia14}. This physical model accounts for relativistic reflection from an incident \texttt{nthcomp} continuum. It can describe both the illuminating and reflected emission. The key parameters include the photon index ($\Gamma$), the accretion disk ionization parameter ($\log \xi$), disk density (log $N$), inclination angle ($i$), inner and outer disk radii, reflection fraction (refl$_{\rm frac}$) and iron abundance relative to solar ($A_{\rm Fe}$). We adopted a single emissivity profile with $q=3$ \citep{Cackett2010, Wilkins12} and fixed the outer disk radius at $R_{\rm out} = 1000 R_G$, where $R_{\rm G}$ is the gravitational radius, since the fit is not sensitive to larger values. The dimensionless spin parameter $a$ was fixed at 0.1885, derived from the spin frequency $\nu = 401$ Hz using the relation $a = 0.47/P_{\rm ms}$, where $P_{\rm ms}$ is the spin period in milliseconds \citep{Braje00}.

\begin{figure*}
\centering
 \includegraphics[width=0.48\linewidth]{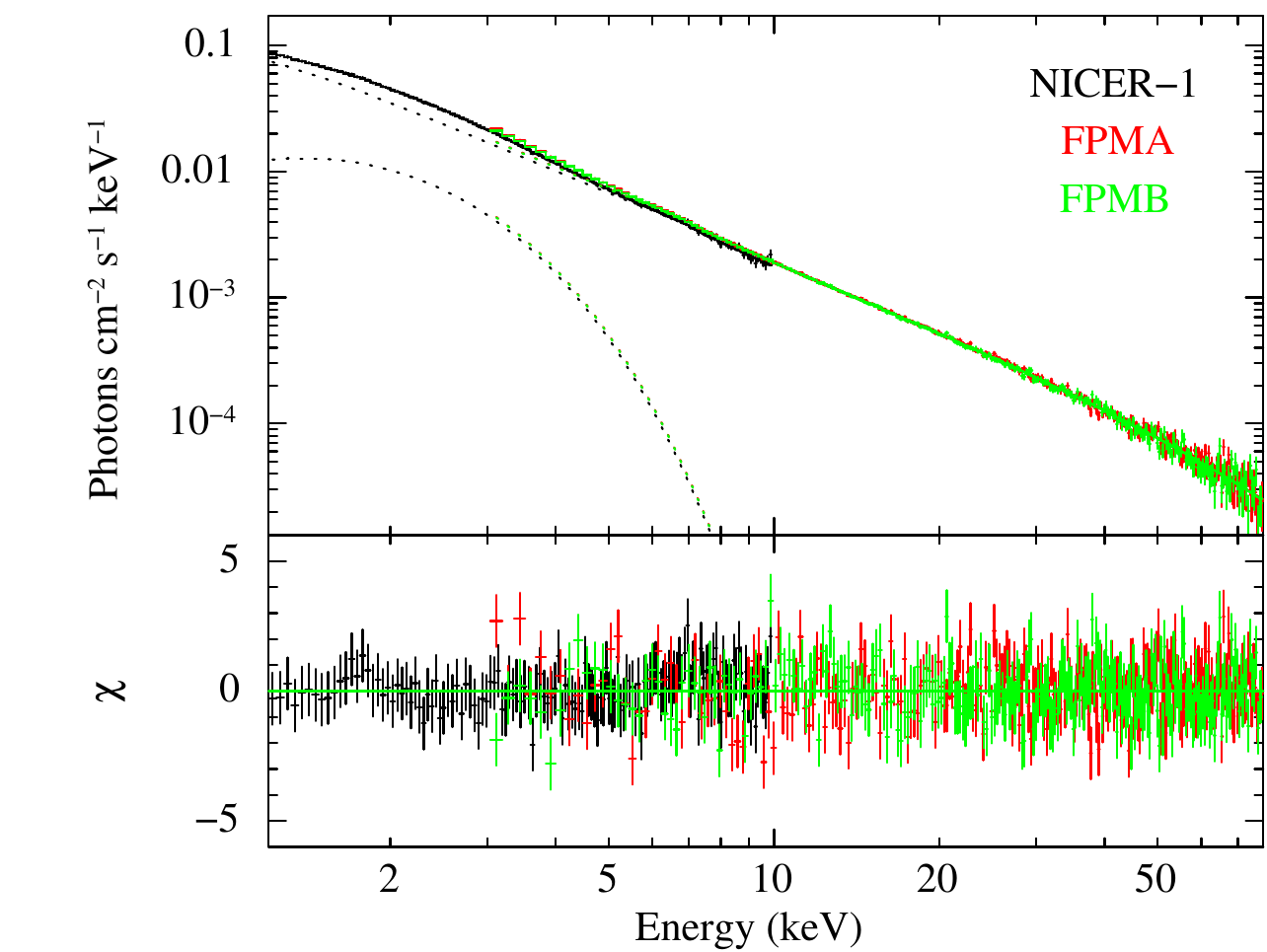}
  \includegraphics[width=0.48\linewidth]{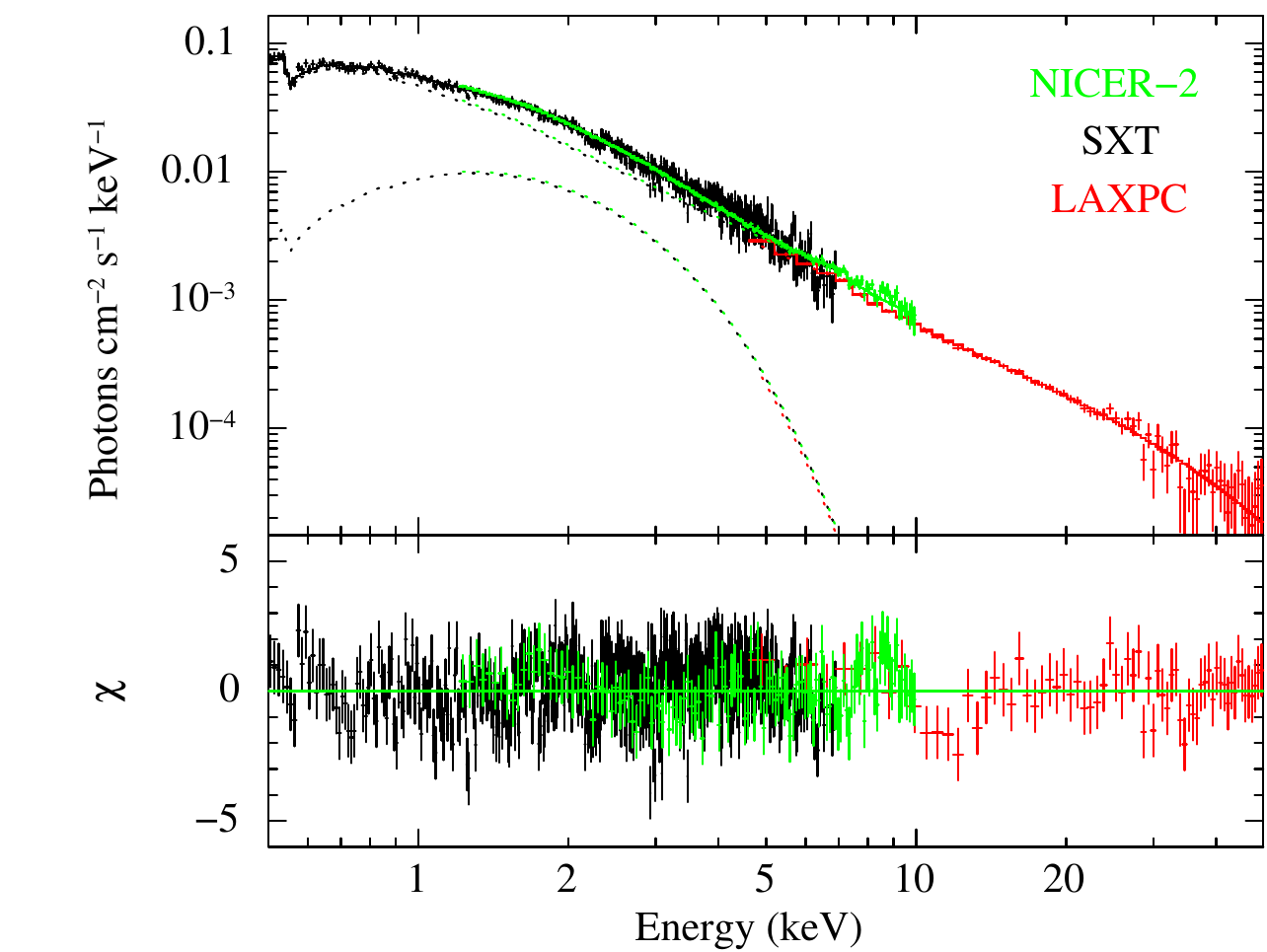}
 \caption{Best-fit broadband energy spectra from Epoch 1 (left) and 2 (right) using \nicer+\nustar\ and \nicer+SXT+LAXPC, respectively, fitted with the combined model \texttt{tbabs*(bbodyrad + relxillCP)}. The lower panels show residuals from the best-fit model.}
\label{fig:spec}
\end{figure*}

The \nicer+\nustar\ spectrum was self-consistently described using the \texttt{tbabs*(bbodyrad+rexillCP)} model. This model yielded a $\chi^2$/dof=592.5/610, indicating a good fit and accounting for the residuals in the 6--7 keV band. The best-fit spectrum is shown in the left panel of Fig. \ref{fig:spec} and the corresponding spectral parameters are listed in Table \ref{tab:fitstat}. During the fitting process, however, we encounter a degeneracy where another local minimum was found corresponding to a truncated inner disk radius at $12 ~R_{\rm ISCO}$ and a nearly neutral reflector ($\log \xi \lesssim 0$). To avoid this degeneracy, we fixed the inner disk radius at the ISCO ($R_{\rm in} = R_{\rm ISCO}$). The preferred solution suggests moderately ionized reflection ($\log \xi \sim 3.4$), an inclination of $31^\circ$, and a poorly constrained iron abundance, with only a lower limit of $A_{\rm Fe} \gtrsim 2$, consistent with previous findings during the 2015 outburst \citep{Salvo2019}.

The same model also adequately describes the Epoch 2 spectrum (\nicer+\astrosat; Table \ref{tab:fitstat}, right panel of Fig.~\ref{fig:spec}). The spectral parameters show notable differences between Epoch 1 and Epoch 2. Between the two epochs, the blackbody temperature remained nearly constant at $\sim$0.6 keV, but its normalization decreased, implying an apparent reduction in the emission region. The Comptonized continuum softened, with the photon index increasing from $\Gamma \sim$1.88 in Epoch 1 to $\Gamma \sim$1.95 in Epoch 2, and the electron temperature dropping from $\sim$32 to $\sim$15 keV, consistent with enhanced Compton cooling during the decay phase.

The relativistic reflection parameters also evolved. The inclination was found to be $\sim 48^\circ$ in Epoch 2 compared to $\sim31^\circ$ in Epoch 1, though this may reflect modeling degeneracies. Previous studies have suggested even higher inclinations ($>$$50^\circ$) \citep{Cackett2009, Papitto2009, Salvo2019}. The inner disk radius remained unconstrained, with an upper limit of $R_{\rm in} < 2.2 R_{\rm ISCO}$. The ionization parameter decreased significantly from $\log \xi \sim 3.4$ to $\log \xi \sim 1.8$, and the iron abundance in Epoch 2 was also poorly constrained, with only a lower limit of $A_{\rm Fe} > 4.6$.

\begin{figure}
\centering
 \includegraphics[width=\columnwidth]{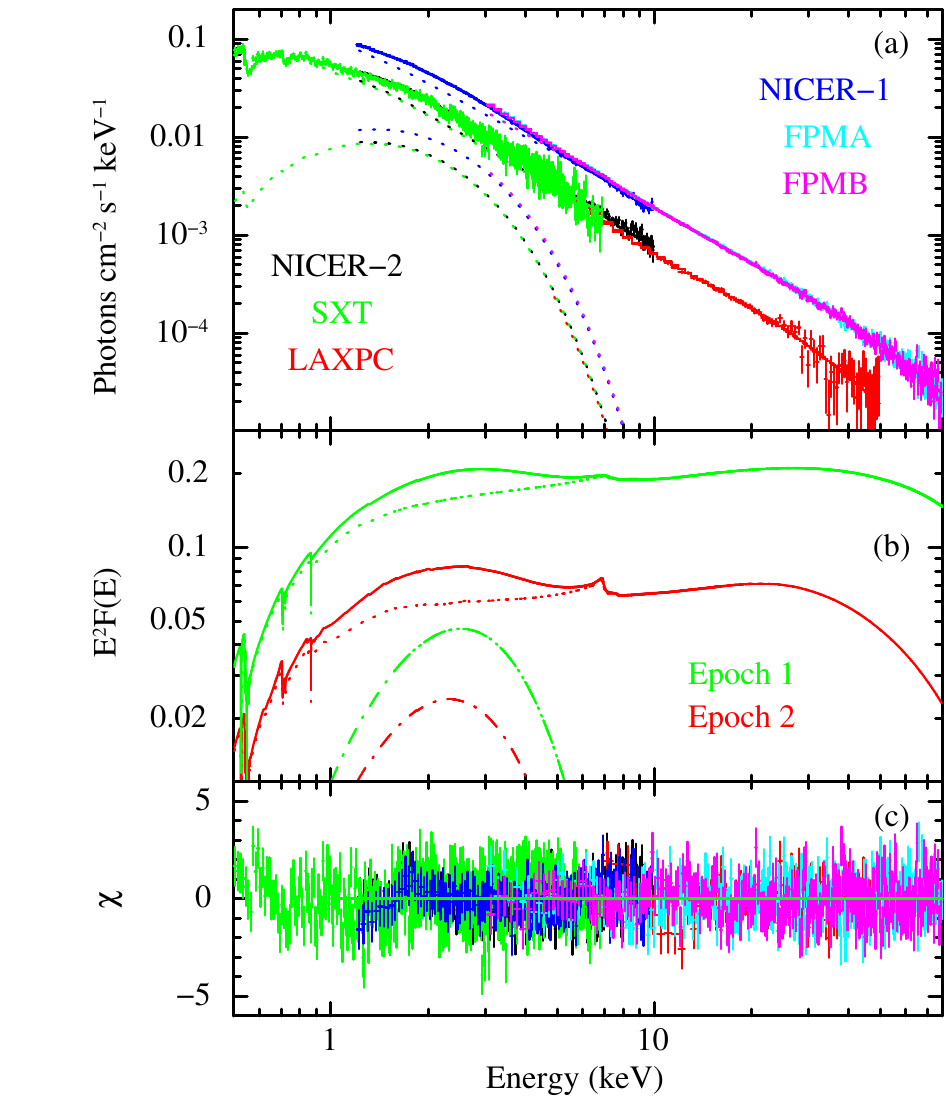}
 \caption{Joint broadband spectral fit of \src\ using \nustar\ (Epoch 1) and \astrosat\ (Epoch 2), together with simultaneous \nicer\ coverage. The spectra are modeled with \texttt{tbabs*(bbodyard + relxillCP)}. (a) Observed spectra with the best-fit model. (b) Comparison of the emission model for both epochs, showing the contributions from the blackbody emission (dashed–dotted lines) and the \texttt{relxillCP} component (dotted lines). (c) Residuals with respect to the best-fit model.}
\label{fig:joint}
\end{figure}

Finally, we performed a joint spectral fit of Epochs 1 and 2 by linking parameters expected to remain constant across epochs. We linked the absorption column density, inclination, iron abundance, and disk density. The resulting best-fit spectra and model components are shown in Fig. \ref{fig:joint}, and the best-fit spectral parameters are reported in Table \ref{tab:fitstat}.
The joint fit yields results consistent with those from the individual fits, with $N_{\rm H} = 2.4 \times 10^{21}$ cm$^{-2}$. The continuum softens as the outburst evolves, with $\Gamma$ increasing from $\sim$1.88 to $\sim$1.99 and the electron temperature decreasing from $\sim$31 to $\sim$18 keV. The blackbody temperature exhibits a mild decrease, accompanied by a reduction in normalization. The reflection modeling suggests a common inclination of $\sim$39$^\circ$ and an iron abundance of $A_{\rm Fe} \sim 3.3$. The normalization of the reflection component also decreases, consistent with the overall decline in flux. The refl$_{\rm frac}$, which is defined as the ratio of the intensity illuminating the disk to the intensity reaching the observer or infinity \citep{Dauser2016}, increased from $\sim$0.1 to $\sim$0.45.
The unabsorbed flux drops from $\sim2 \times 10^{-9}$ to $\sim7 \times 10^{-10}$ \erg, corresponding to a luminosity decrease from $2.8 \times 10^{36}$ to $1 \times 10^{36}$ \lum\ (assuming a distance of 3.5 kpc; \citealt{Galloway05}).

\renewcommand{\arraystretch}{1.2}
\begin{table*}
	\centering
	\caption{Best-fit spectral parameters of \src{} with spectral model \texttt{tbabs*(bbodyrad + relxillCP)}.}
	\label{tab:fitstat}
	\resizebox{0.8\linewidth}{!}{
	\begin{tabular}{l|cc|cc} 
		\hline
    Parameters & Epoch 1 & Epoch 2 & \multicolumn{2}{c}{Joint Fit}  \\
   & \nicer+\nustar & \nicer+\astrosat & Epoch 1 & Epoch 2  \\
		\hline
$N_{\rm H}$ ($10^{21}$ cm$^{-2}$) & $2.9^{+0.6}_{-0.2}$ & $2.2 \pm 0.1$ & \multicolumn{2}{c}{$2.41 \pm 0.15$}  \\

$kT_{\rm BB}$ (keV) & $0.60 \pm 0.01$ & $0.57 \pm 0.01$ & $0.63 \pm 0.01$ & $0.576 \pm 0.014$   \\  
        Norm & $77 \pm 9$ & $52 \pm 7$ & $63 \pm 6$ & $46 \pm 6$ \\
    $R_{\rm BB}$ (km)$^{a}$ & $3.1 \pm 0.2$ & $2.5 \pm 0.2$ & $2.8 \pm 0.1$ & $2.4 \pm 0.2$ \\  [1ex]
                   
Incl (deg) & $31.3^{+3.6}_{-6.4}$ & $48.3^{+5.7}_{-2.1}$ & \multicolumn{2}{c}{$39.3^{+3.6}_{-1.9}$}  \\ 
        $R_{\rm in}$ ($R_{\rm ISCO}$) & $1^{\rm fixed}$ & $<2.2$ & $1^{\rm fixed}$ & $<3.1$\\  
        
        $\Gamma$ & $1.88 \pm 0.01$ & $1.95^{+0.02}_{-0.01}$  & $1.88 \pm 0.01$ & $1.99 \pm 0.03$ \\
        
        log $\xi$ & $3.4^{+0.2}_{-0.3}$ & $1.87^{+0.24}_{-0.13}$ & $3.4 \pm 0.2$ & $1.85^{+0.45}_{-0.15}$\\ 
        
        log ($N$ cm$^{-3}$) & $18.0_{\rm pegged}^{+1.6}$ & $16.9^{+0.3}_{\rm pegged}$ & \multicolumn{2}{c}{$17.0^{+0.8}_{\rm pegged}$} \\
        
        $A_{\rm Fe}$ & $>2.0$ & $>4.6$ & \multicolumn{2}{c}{$3.3 \pm 1.1$}  \\ 
        
        $kT_{\rm e}$ (keV) & $32.4^{+5.5}_{-3.5}$ & $15.2^{+3.4}_{-2.5}$ & $31.3^{+3.9}_{-2.9}$ & $18.2^{+8.0}_{-3.5}$ \\
        
        refl$_{\rm frac}$ & $8.8^{+3.5}_{-2.7}$ $\times10^{-2}$ & $0.51^{+0.13}_{-0.11}$ & $0.10^{+0.04}_{-0.01}$ & $0.45^{+0.12}_{-0.16}$ \\
        
        Norm ($10^{-4}$) & $21.8 \pm 0.8$ & $7.2 \pm 0.3$ & $21.6 \pm 0.9$ & $7.5 \pm 0.4$\\[1ex]
                          
        $C_{\rm FPMA/LAXPC}$ & $1$ (fixed) & $1$ (fixed) & $1$ (fixed) & $1$ (fixed) \\
        $C_{\rm NICER}$ & $0.925 \pm 0.004$ & $1.13 \pm 0.02$ & $0.927 \pm 0.005$ & $1.13 \pm 0.02$ \\
        $C_{\rm FPMB/SXT}$ & $0.992 \pm 0.002$ & $1.16 \pm 0.02$ & $0.992 \pm 0.003$ & $1.16 \pm 0.02$\\[1ex]
                          
        Flux$^{b}_{\rm 0.1-100~keV}$ & $1.96 \times 10^{-9}$ & $6.9 \times 10^{-10}$ & $ 1.94 \times 10^{-9}$ & $7.2 \times 10^{-10}$ \\  

        L$_{\rm 0.1-100~keV}^{c}$ & $2.87 \times 10^{36}$ & $1.01 \times 10^{36}$ & $ 2.84 \times 10^{36}$ & $1.05 \times 10^{36}$ \\                     
\hline                  
        $\chi^2$/dof & 592.5/610 & 749.8/679 & \multicolumn{2}{c}{1257/1293} \\                    
		\hline
        \multicolumn{5}{l}{\textbf{Notes:} All errors and upper limits reported in this table are at a 90\% confidence level ($\Delta \chi^2=2.7$).}\\
        \multicolumn{5}{l}{$^{a}$The emission radius is calculated assuming a distance of 3.5 kpc.}\\
		\multicolumn{5}{l}{$^b$ Unabsorbed flux in units of \erg.}\\
  		\multicolumn{5}{l}{$^c$X-ray Luminosity in units of \lum.}\\
        
	\end{tabular}}
\end{table*}

\section{Discussion}
\label{dis}
\href{}{}
\src\ was observed in its tenth known outburst in 2022. We report results from our broadband timing and spectral study of \src\ based on \nustar\ and \astrosat\ observations performed near the peak (Epoch 1) and decay phase (Epoch 2) of outburst, respectively, simultaneously with \nicer. Our joint analysis of \nicer, \nustar, and \astrosat\ observations provides a comprehensive insight into the evolution of the pulse properties and broadband spectrum over these two distinct stages.

\subsection{Timing properties}
 
The coherent pulsations at $\sim$401 Hz were detected throughout both epochs, confirming the persistent channeling of accretion flow onto the magnetic poles during the outburst. Pulse timing results obtained are compatible within the errors with the solution obtained with \nicer\ \citep{Illiano23}. The pulse profiles in the energy band 3--30 keV from both epochs are well described by two harmonically related sinusoidal components, with the fundamental amplitude increasing from $\sim$4\% near the outburst peak to $\sim$6\% during the decay. Notably, during the reflaring stage of the 2022 outburst, the fundamental amplitude was reported to reach even higher values, up to $\sim$10\% during the low-flux state \citep{Ballocco25}. This increase likely reflects changes in the accretion column geometry or hotspot visibility as the accretion rate declines \citep{Kajava2011, Markozov2024}. 
The harmonic content remains broadly similar between epochs in the hard X-ray band, although a weak third harmonic component may be present, albeit with low significance.

\subsection{Energy Dependence of Pulsed emission}

The fractional amplitude and phase lag were found to be energy-dependent during this outburst, similar to previous outbursts \citep[e.g.,][]{Hartman2009, Sanna2017, Bult2020, Sharma2023-SAX}. We checked for the energy dependence in the 0.5--30 keV range. During both epochs, the fundamental amplitude increases with energy up to $\sim$3 keV before decreasing at higher energies. The peak fractional amplitude reached $\sim$7\% during Epoch 1, while it peaked around 9\% for Epoch 2. Similar trends were also observed with XMM-Newton observation taken during the reflaring phase \citep{Ballocco25}.
The harmonic component in Epoch 1 increased steadily with energy, reaching $\sim$2\% at the higher energies, while the harmonic became more prominent above 5 keV in Epoch 2, reaching $\sim$3\% at higher energies. A similar trend of increase in the fractional amplitude of the harmonic with energy has been observed in the previous outburst of 2019 \citep{Bult2020, Sharma2023-SAX}. However, the energy dependence of the fractional amplitudes varies considerably between different outbursts and luminosities \citep[e.g.,][]{Hartman2009, Patruno2009, Sanna2017, Bult2020}. During Epoch 1, a dip feature in the fractional amplitude around the 6--7 keV energy range (corresponding to the Fe line) is evident, which is absent from Epoch 2. A similar drop in fractional amplitude near the Fe line was previously observed during the 2008 and 2015 outbursts \citep{Patruno2009, Sanna2017}, whereas no such feature was seen in the 2019 outburst \citep{Sharma2023-SAX}.

The time lags during the 2022 outburst exhibited the typical behavior seen in \src. The hard photons arrive earlier than soft ones, producing a negative lag that increases with energy \citep[e.g.,][]{Cui1998, Gierlinski2002, Hartman2009, Ibragimov2009, Sharma2023-SAX}. In both epochs, the fundamental component exhibited soft lags reaching 0.2--0.3 ms. 
The harmonic component showed a similar order of lag in Epoch 2, while no significant lag was observed during Epoch 1. The time lags in \src\ are known to depend on source flux, typically becoming more pronounced as the accretion rate decreases. Soft lags in the fundamental component are observed throughout the outburst's evolution, while lags for harmonics are less pronounced or become hard lags at higher luminosities \citep{Hartman2009, Ibragimov2009}. 
During the 2019 outburst, both the fundamental and harmonic components showed lags of comparable magnitude but with the opposite trend when observed near the outburst peak with \astrosat\ \citep{Sharma2023-SAX}. These lags can be explained by the model where the soft thermal emission and Comptonization emissivity (or beaming) patterns are different, each affected differently by rapid stellar rotation \citep{Gierlinski2002, Poutanen2003, Ibragimov2009}. The stronger pulsation amplitude and persistent negative lags in Epoch 2 suggest a reduction in scattering depth and a more compact emission geometry as the outburst decayed \citep{Poutanen2004, Viironen2004, Falanga2007, Mushtukov2023}.

\subsection{Spectral properties}

The broadband X-ray spectrum of \src\ in both epochs is well described by the combination of a soft thermal component, a Comptonized continuum, and a significant relativistic reflection component from the accretion disk. A joint fit to the spectra from both epochs, with linked parameters expected to remain unchanged, reinforces the trends inferred from the individual fits. A common inclination ($\sim$39$^\circ$) and iron abundance ($A_{\rm Fe}\sim3.3$) were observed. We observed pronounced spectral evolution between the two epochs. The X-ray luminosity decreases by a factor of $\sim$3 from Epoch 1 to Epoch 2, reflecting the declining accretion rate. The Comptonized continuum softens, with the photon index increasing from $\Gamma \sim 1.88$ to $\Gamma \sim 1.99$, while the electron temperature decreases from $\sim$31 to $\sim$18 keV. These changes indicate enhanced Compton cooling as the accretion rate falls. The ionization parameter drops from $\log\xi \sim 3.4$ to $\log\xi \sim 1.8$, and the reflection fraction increases, suggesting that as the disk cools and becomes less ionized, a larger fraction of the Comptonized flux is intercepted and reprocessed by the disk \citep{Petrucci2001, Garcia2010}. 

Although spectral softening is evident between Epochs 1 and 2, the best-fitting spectral parameters indicate that \src\ remained in the hard spectral state during both epochs. The hardness ratio measured with \nustar\ and \astrosat\ showed a significant decrease between the two epochs, while the \nicer\ data reveal only a mild downward trend. This reduction in hardness is consistent with the spectral softening inferred from our broadband fits, likely reflecting changes in the Comptonizing plasma and disk emission as the outburst evolves \citep{Poutanen2003, Falanga2005, Ferrigno2017, Patruno2021}. Despite this evolution, no evidence for a transition to a soft state is observed in the 2022 outburst, similar to the behavior reported during the 2019 event. In contrast, during the 2015 outburst, \citet{Salvo2019} reported the state transition in \src\ with XMM-Newton (soft state) and \nustar\ (hard state) observations.  

\subsection{Comparison with Other Studies of the 2022 Outburst}

Several studies have reported results from the 2022 outburst of \src\ using different instrumental combinations and scientific focuses. In particular, \citet{Kaushik25} and \citet{Bruce26} investigated the broadband spectral properties using NICER in combination with hard X-ray observations, while \citet{Ballocco25} focused on the temporal and spectral behavior during the reflaring phase using XMM-Newton and HST data. In addition, \citet{Dorsman26} modeled the pulse profiles from \nicer\ to constrain the emission geometry. While these works employ broadly similar spectral components, our analysis provides complementary and, in several respects, new insights into the 2022 outburst. 

The main differences and advances of the present work can be summarized as follows:
\begin{itemize}
    \item Our analysis exploits strictly simultaneous broadband coverage, combining \nicer\ with \nustar\ (Epoch~1) and \astrosat\ (Epoch~2), and explicitly tracks the evolution of both pulsed and spectral properties between the peak and decay phases of the outburst.

    \item Our work also extends previous work by incorporating a detailed timing analysis and energy-resolved pulse properties. For the first time for the 2022 outburst, we report the energy dependence of the pulsed fractional amplitude and phase lags across two distinct outburst phases: the peak and decay phases. The detection of epoch-dependent changes in harmonic content, the presence of a dip in the fractional amplitude of the fundamental at the Fe-line energy during the peak, and the evolution of soft lags provide new constraints on the geometry of the emission regions that were not addressed in previous studies.
       
    \item \citet{Bruce26} performed broadband spectral modeling using a single NICER observation closest in time to the corresponding \nustar\ exposure. In contrast, we explicitly merged multiple \nicer\ observations to construct event files that strictly overlap the \nustar\ and \astrosat\ time windows. This approach captures intrinsic flux variability within each broadband observation and yields a more representative description of the instantaneous spectral state, which is particularly important given the rapid evolution in luminosity observed during the 2022 outburst.
        
    \item From a spectral perspective, our joint fitting of Epochs~1 and~2 allows us to track the intraoutburst evolution of spectral parameters in a self-consistent manner. While \citet{Bruce26} constrained reflection parameters by jointly fitting the 2019 and 2022 outbursts, our analysis focuses specifically on the evolution within the 2022 outburst itself. Our results suggest a dynamically changing corona–disk coupling on timescales of days, complementing the longer-term comparisons presented in earlier works.
    Although broadly similar continuum and relativistic reflection models are employed, differences are observed between our results and those of \citet{Bruce26}, particularly in the inferred inclination, disk ionization, and reflection fraction. These discrepancies are likely attributable to differences in modeling approaches. For instance, we allowed the iron abundance to vary freely, finding values higher than solar, and we did not model the $\sim$1~keV feature, which can influence the reflection continuum at low energies.

    \item \citet{Kaushik25} classified the source as being in a soft spectral state throughout the 2022 outburst. This interpretation differs from our results, which indicate that \src\ remained in the hard state during both epochs. Our conclusion is based on broadband spectral modeling extending up to $\sim$50--78 keV, which provides stronger constraints on the Comptonized continuum and high-energy cutoff, and is consistent with the findings of \citet{Bruce26} and \citet{Ballocco25}. The discrepancy likely arises from differences in the spectral approach and energy coverage, as \citet{Kaushik25} based their classification primarily on modeling of the \nicer\ data. In addition, \citet{Kaushik25} reported fractional rms amplitude of red noise components in the range of $\sim$12--26\%, which are characteristic of NS-LMXBs in the hard state, where rms values of $\gtrsim$10--20\% are typically observed \citep[see, e.g.,][]{Munoz, Sharma2023-SAX, Sharma25XB}.
   
    \item In other studies of the 2022 outburst, the $\sim$1~keV feature detected in \nicer\ spectra was modeled using an additional Gaussian component \citep{Kaushik25, Bruce26, Dorsman26}, as similar features were reported during earlier outbursts \citep[e.g.,][]{Sharma2023-SAX}. However, during the 2022 outburst, we did not include this feature in our spectral modeling, as it is not detected with \sxt\ and may be affected by instrumental background effects or uncertainties in the \nicer\ effective area \citep[see, e.g.,][]{Hall25}. Moreover, if this feature were intrinsic to the source and originated from the same region and/or ionization state as the Fe~K emission, it would be expected to be reproduced by a self-consistent reflection model, which is not observed \citep{Bruce26}.

    \item The presence of a soft thermal component with a temperature of $\sim$0.6~keV and an emitting radius of $\sim$2--3~km is broadly consistent with previous findings. \citet{Ballocco25} reported a similar size of emitting region during the reflaring phase, albeit with a lower temperature of $\sim$0.36~keV, likely reflecting cooler thermal emission associated with a lower accretion rate.
\end{itemize}

At the time of writing, no refereed publications are available for the 2025 outburst of \src. Preliminary reports suggest comparable peak luminosities with a broadly similar outburst morphology \citep{Ballocco25ATel1, Ballocco25ATel2}, but no detailed broadband timing or reflection-based spectral analysis has yet been presented. Our results from the 2022 outburst, therefore, provide a valuable reference framework for interpreting future observations of the 2025 outburst, particularly in assessing whether similar coronal cooling, evolution of reflection, and pulse-lag behavior recur across successive outbursts.

\section{Conclusions}
\label{summary}

We have presented a detailed timing and broadband spectral investigation of the AMXP \src\ during its 2022 outburst, based on coordinated observations with \nicer, \nustar\ and \astrosat\ covering both the peak and decay phases. Our main findings can be summarized as follows: 

\begin{itemize}
    \item Coherent pulsations at $\sim$401~Hz persist throughout the outburst, with the pulsed fractional amplitude of the fundamental increasing from the peak to the decay phase and evolving harmonic component as the accretion rate declines.   
    \item The energy dependence of the pulse amplitude and time lags exhibits clear epoch-dependent behavior, including a transient dip at the Fe-line energy during the outburst peak.
    \item The broadband spectrum shows pronounced evolution, with spectral softening, a decrease in coronal electron temperature, and a reduction in disk ionization as the accretion rate declines. This behavior is consistent with gradual spectral softening within the hard state rather than a full state transition.
    \item An increasing reflection fraction during the decay phase suggests a more compact corona and enhanced disk covering fraction at lower accretion rates.
    \item The combined timing and spectral results point to a coupled evolution of the corona, accretion disk, and magnetosphere as the system transitions from the peak to the decay phase of the outburst.
\end{itemize}
    
Overall, our analysis demonstrates the importance of strictly simultaneous broadband coverage and joint timing and spectral diagnostics in disentangling the physical processes governing AMXPs. The results place the 2022 outburst of \src\ in the broader context of its recurrent activity and provide a robust reference for forthcoming detailed studies of the 2025 outburst.

\begin{acknowledgments}
This research has made use of \astrosat\ data, obtained from the Indian Space Science Data Centre (ISSDC), and \nustar\ and \nicer\ data obtained from the High Energy Astrophysics Science Archive Research Center (HEASARC). We thank the LAXPC Payload Operation Center (POC) and the SXT POC at TIFR, Mumbai, for providing the necessary software tools. We thank the anonymous referee for constructive comments and suggestions.
\end{acknowledgments}




%
\facilities{NICER, NuSTAR, AstroSat.}

\software{Heasoft, Xspec.}




\section*{Appendix}
\section*{A: Pulse profile shape}
\addcontentsline{toc}{section}{Appendix A: Pulse profile shape}
%

The pulse profiles shown in Fig.~\ref{fig:pp} appear slightly offset from the baseline level of unity and are not symmetrically distributed around it. This effect is more pronounced in the \nustar\ and \astrosat\ profiles. The apparent offset arises from the combined contribution of the fundamental and harmonic components to the total pulse shape. While the individual fundamental and harmonic components are each symmetric about the baseline level of unity, their superposition produces an asymmetric total profile.

\begin{figure}[h]
\centering
 \includegraphics[width=0.99\columnwidth]{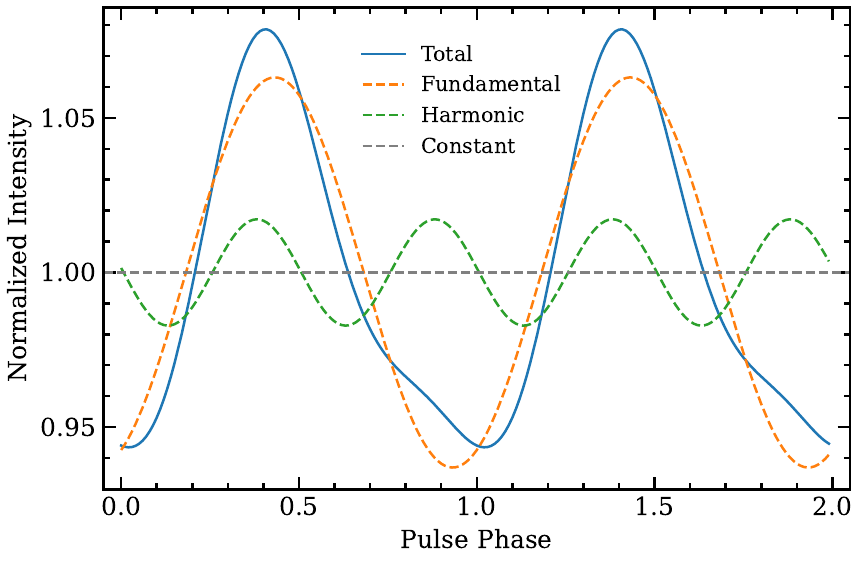}
 \caption{Decomposition of the LAXPC pulse profile into its fundamental and harmonic components. The individual components are symmetric about the baseline level of unity, while their superposition produces an asymmetric total pulse profile due to their relative phase alignment.}
\label{fig:check}
\end{figure}

Fig.~\ref{fig:check} illustrates this effect by showing the individual contributions of the fundamental and harmonic components, along with their combined pulse profile. In particular, the harmonic component contributes additional flux near the pulse maximum as well as at other pulse phases, leading to an enhanced peak and a shallower minimum. As a result, the total pulse profile spans an asymmetric range due to the relative phase alignment between the harmonic and fundamental components.

\bibliography{ref}{}
\bibliographystyle{aasjournalv7}



\end{document}